\newcommand{\ket}[1]{\vert #1 \rangle}
\newcommand{\bra}[1]{\langle #1 \vert}

\newcommand{\braket}[2]{\langle #1 \vert #2 \rangle}

\newcommand{\ketbra}[2]{\vert #1 \rangle  \langle #2 \vert}

\newcounter{myctr}
\def\myitem{\refstepcounter{myctr}\bibfont\noindent\ifnum\themyctr>9\else\phantom{0}\fi\hangindent17pt\themyctr.\enskip}

\documentclass{ws-ijqi}
\usepackage{hyperref}
\usepackage[super,sort,compress]{cite}
\usepackage{xcolor}
\usepackage{url}
\begin{document}
	
	\catchline{}{}{}{}{}
	
	\title{QUANTUM NON-GAUSSIANITY FROM AN INDEFNITE CAUSAL ORDER OF GAUSSIAN OPERATIONS}
	
	\author{SEID KOUDIA}
	
	\address{Laboratoire de Physique Th\'eorique, Facult\'e des Sciences Exactes, Universit\'e de Bejaia\\
		06000 Bejaia, Algeria\\
		seid.koudia@univ-bejaia.dz}
	
	\author{ABDELHAKIM GHARBI}
	
	\address{Laboratoire de Physique Th\'eorique, Facult\'e des Sciences Exactes, Universit\'e de Bejaia\\
		06000 Bejaia, Algeria\\
		abdelhakim.gharbi@univ-bejaia.dz}

	\maketitle
	
	\begin{history}
	\end{history}
	
	\begin{abstract}
		Quantum Non-Gaussian states are considered as a useful resource for many tasks in quantum information processing, from quantum metrology and quantum sensing to quantum communication and quantum key distribution. Another useful tool that is growing attention is the newly constructed quantum switch. Its applications in many tasks in quantum information have been proved to be outperforming many existing schemes in quantum communication and quantum thermometry. In this contribution, we are addressing this later to be very useful to engineer highly non-Gaussian states from Gaussian operations whose order is controlled by degrees of freedom of a control qubit. The non-convexity of the set of Gaussian states and the set of Gaussian operations guarantees the emergence of non-Gaussianity after postselection on the control qubit deterministically, in contrast to existing protocols in the literature. The non-classicality of the resulting states is discussed accordingly.  
		
	\end{abstract}
	
\section{\label{sec:1}Introduction}

Continuous variable quantum information plays a pivotal role in the rapidly evolving field of quantum technologies. It uses continuous variable systems (CV) that are described by canonically conjugated variables, position and momentum, represented on an infinite dimensional Hilbert space \cite{Braunstein_2005, andersen2010continuous, Adesso_2014, aless2005gaussian}. 
\par
In particular, Gaussian states and Gaussian operations played important roles on CV quantum information processing. In addition of being readily accessible in quantum optical experiments and easy to implement, they proved useful in many quantum protocols from quantum teleportation and quantum key distribution, to quantum enhanced-measurements. Although, this class of CV systems faces many limiting no-go theorems in their applicability power in CV quantum information processing,  to ensure an exponential speedup quantum computing \cite{pfister2019continuous, efficient}. 
\par
However, Non-Gaussian states and non-Gaussian operations guarantee this task without suffering from fundamental impossibilities regarding fault tolerance \cite{fault} and quantum error correction \cite{nongo}. They proved to be important for many other foundational questions and applications such as loophole-free violation of Bell inequalities \cite{loophole, loophole2}, optimal cloning \cite{cloning}, CV measurement-based quantum computing \cite{computing, computing1}. Notably, non-Gaussian states such as  NOON states \cite{NOON}, Schrödinger cat states \cite{cat}, and non-Gaussian operations such as photon-number addition \cite{addition} and photon-number subtraction \cite{subtraction}, can improve the quality of entanglement and correspondingly tasks such as teleportation \cite{teleportation}.\\
This class of CV systems has got accordingly, a remarkable attention in the last decade. They are theoretically analyzed and experimentally realised. On one side, increased scrutiny has been given to their characterisation and quantification in a resource theoretic framework \cite{shapiro, Albarelli_2018}. On the other side, special focus is dedicated to the engineering of non-Gaussian states. This later faces many problems, for the procedure suffers from the low efficiency due to the small values of susceptibilities of the nonlinear media that are required for the generation of non-classical light, in addition to the non-negligible noise added when the signal undergoes amplification, preserving the uncertainty relation \cite{caves}.
\par
Recently, many attempts to foil these issues have been suggested, and they all rely on conditional generation of non-Gaussian states by measurement protocols with post-selection on a given outcome, such as photon addition and photon subtraction, which makes them non-deterministic. They only achieve the generation of non-Gaussian states with some given probability of success \cite{ralph, Hamza}.
\par
From another side, a growing interest in quantum causality is witnessed \cite{pienaar2018quantum, barrett2019quantum, common, Costa_2016}, due to the recently introduced notion of indefinite causal ordering, which is a result of an extension of quantum mechanics to include higher order processes. It has been known in the framework of quantum combs, and later in the process matrix formalism, which allows for the possibility of a coherent superposition of orders of quantum operations. A simple realisable process that entails such a superposition is the quantum switch \cite{chirib,1}, where the order of processes acting on a given system becomes entangled with a quantum degree of freedom. This new type of quantum correlations, i.e. \textit{causal-nonseparability} \cite{causal, Branciard_2016}, is a resource that provides many advantages in different tasks \cite{switch2019, salek2018quantum, switch1, Giulia,switch}. Many experiments and implementations of the Switch were devised \cite{2,Rubino_2017,Goswami_2020,3}, in different contexts, ranging from quantum gravitational control of temporal order \cite{Zych_2019,zych2018relativity}, to quantum communication and computation advantages \cite{giuliarub,4}. 

\par
In this paper, we address the generation of non-Gaussian states by use of an indefinite causal order between Gaussian operations, namely, displacement and squeezing operators, acting on an initially Gaussian state which is taken to be the single mode vacuum state. The indefinite relative order in which the Gaussian operations are acting on the target state, is realized by the aid of a control qubit degree of freedom, that is allowed to evolve together with the target in a Quantum switch. At the end, a measurement on the control qubit is performed in the coherent basis, and accordingly, the indefinite order between Gaussian operations is achieved.  Clearly, this superposition of causal orders of operations in the switch is obtained by postselection on one of the measurement outcomes on the control qubit. The properties of the outcoming conditional postselected states will be studied. Namely, non-Gaussianity in terms of the quantum relative entropy with reference to a Gaussian state \cite{Genoni_2008}, and non-classicality by means of the negativity of the Wigner's function \cite{negativity}, will be both explored. In contrast to existing protocols,  we will prove that the adopted scheme using the resource of causal-non-separability, allows for the deterministic generation of highly non-classical and non-Gaussian states, depending on the squeezing and the displacement parameters, in the sense that post-selection on both outcomes of the measurement performed on the control qubit will generate equally a non-Gaussian and non-classical state, in contrast to existing protocols, which allows for the emergence of the desired states with a low probability of success.
\par
The paper is organised as follows. In Sec.~(\ref{sec:2}) we briefly give some properties of Continuous variable systems. Furthermore, we present useful tools for the characterization of non-Gaussianity and non-classicality in quantum states. In particular, we will focus on the quantum relative entropy of non-Gaussianity and Wigner's negativity as quantifiers. In sec.~(\ref{sec 3}), we give a summary of our results. We present how causal-nonseparability, in particular superposition of causal orders of Gaussian operations in the quantum switch, can be useful for state generation. The properties of relevant interest of the resulting conditional states from the proposed control protocol will be treated. Specifically, the non-Gaussianity and non-classicality features will be discussed. Accordingly, we will show that both conditional states present these features, thus, the deterministic engineering of non-Gaussian and non-classical states from an indefinite causal order of Gaussian operations. Finally, Sec.~(\ref{conc}) closes the paper with concluding remarks.

\section{\label{sec:2}Preliminaries}

In this section we set some notations and discuss briefly basic properties of CV systems. Properties characterizing non-Gaussian states and the quantification of their non-Gaussianity and non-classicality will be presented. For instance, we will define the quantum relative entropy based measure for non-Gaussianity and the Wigner's negativity based measure for non-classicality which we are adopting in the following sections. We give a brief overview as well on the superposition of causal orders in the quantum switch, without entering into the details of quantum causal modelling in the framework of the process matrix formalism that underlies its structure.

\subsection{\label{CV}Continuous variable systems} 
 A single mode CV system is represented by annihilation and  creation operators which satisfy the commutation relations $\left[\hat{a},\hat{a}^{\dagger}\right]=1$, $\left[\hat{a},\hat{a}\right]=0=\left[\hat{a}^{\dagger},\hat{a}^{\dagger}\right]$. These operators can be written in terms of quadrature operators $\hat{\textbf{Q}}=(\hat{q},\hat{p})^T$, which satisfy the canonical commutation relation $\left[\hat{q},\hat{p}\right]=i$ for $\hat{a}=\hat{q}+i\hat{p}$ and $\hat{a}^\dagger=\hat{q}-i\hat{p}$.
\par
A quantum state $\hat{\rho}$ can be described by its Wigner's characteristic function $\chi(\alpha,\hat{\rho})=\hbox{Tr}\left[\hat{\rho}\hat{D}(\alpha)\right]$, where $D(\alpha)=e^{\alpha\hat{a}^{\dagger}-\alpha^*\hat{a}}$ is the displacement operator, and $\alpha=x+iy$. The Wigner's function of the state $\hat{\rho}$ is defined to be the Fourier transform of the Wigner's characteristic function, and it is given for Gaussian states by 

\begin{equation}
    W_{\rho}(\textbf{Q})=\frac{\hbox{exp}\left[-\frac{1}{2}(\textbf{Q}-\langle\hat{\textbf{Q}}\rangle_{\rho})^T\textbf{V}^{-1}(\textbf{Q}-\langle\hat{\textbf{Q}}\rangle_{\rho})\right]}{2\pi\sqrt{\textbf{V}}}\label{eq: wigner gaussian}
\end{equation}
where $\textbf{V}$ is the covariance matrix which is given by 
\begin{equation}
    \textbf{V}_{ij}=\frac{1}{2}\langle\{\hat{Q}_i,\hat{Q}_j\}\rangle_{\rho}-\langle \hat{Q}_i\rangle_{\rho} \langle \hat{Q}_j\rangle_{\rho}\label{eq: covariance}
\end{equation}
 Gaussian states are completely characterized by their first and second moment given by the covariance matrix. 
 
 \subsection{\label{nongaussianity} Quantifying non-Gaussianity}
 Following the resource theory of non-Gaussianity, there are two approaches: Resource theory for non-Gaussian states and the resource theory for non-Gaussian operations. The first only characterizes the non-Gaussianity of quantum states, whereas the former characterizes non-Gaussianity at the level of quantum operations. In this paper we are following the first approach which uses a quantum relative entropy, which is considered as a measure of distinguishability between two states, as  monotone of non-Gaussianity. By exploiting a reference quantum state $\hat{\rho}_G$, the non-Gaussianity measure of a given state $\hat{\rho}$ is given by 
 \begin{equation}
     \delta_{nG}=S(\hat{\rho}\mid\mid\hat{\rho}_G)=S(\hat{\rho}_G)-S(\hat{\rho}) 
 \end{equation}
for $S$ being the von Neumann entropy. The last equality results from the assumption that $\hat{\rho}_G$ has the same covariance matrix $\textbf{V}$ as $\hat{\rho}$.
For pure states $\hat{\rho}$ this measure reduces to 
\begin{equation}
    \delta_{nG}=h(\sqrt{\det\textbf{V}})\label{eq: nG}
\end{equation}
where $h(x)=(x+\frac{1}{2})\log(x+\frac{1}{2})-(x-\frac{1}{2})\log(x-\frac{1}{2})$. This is because the von Neumann entropy of single mode Gaussian states is fully determined by their covariance matrix.
\begin{figure}[t]
    \centering
    \includegraphics[width=0.7\columnwidth]{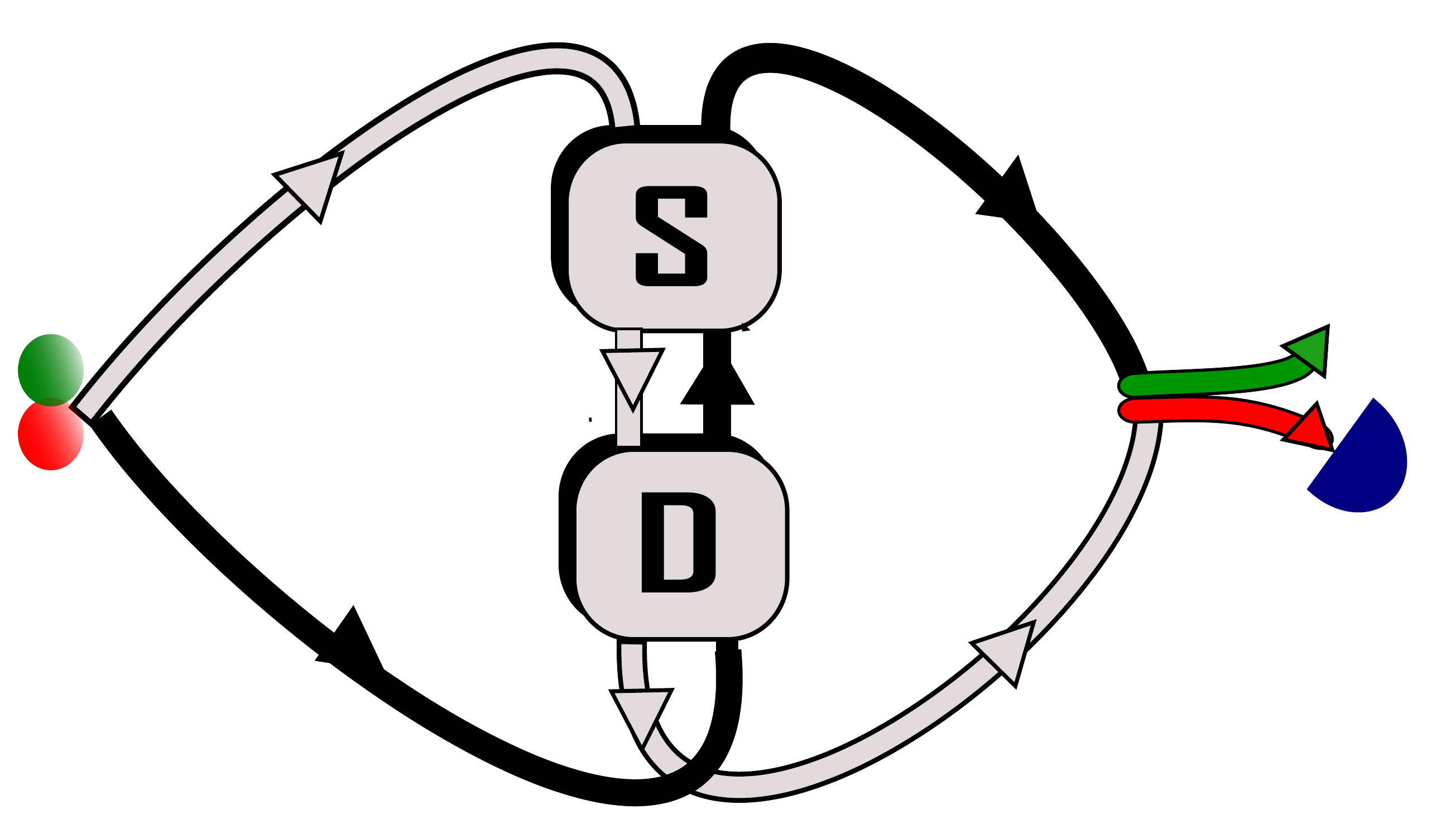}
    \caption{A scheme representing the quantum switch that is implementing an indefinite causal order between squeezing operator $S$ and displacement operator $D$ acting on an initial bosonic vacuum state (green) coupled to a qubit degree of freedom (red), on which a measurement is performed at the end\label{fig:qs}}
\end{figure}
\subsection{\label{sec: nonclassicality}Quantifying non-classicality}
The description of CV states in terms of quasi-probability distributions on classical phase space proved very useful in quantum optics. One advantage of these representations is to characterize the limit between classical and quantum states of light. An indicator for non-classicality of a CV state has been procured from properties of its Wigner function.  It has been shown \cite{negativity} that the volume of the negative part of the Wigner function is a quantifier of the non-classicality of the state. Formally
\begin{equation}
    \delta_{nC}=\int\int\lvert W(q,p)\rvert dqdp-1\label{eq:nc} 
\end{equation}
This non-classicality measure will be adopted in the next section.

\section{Quantum non-Gaussianity from a superposition of causal oders\label{sec 3}}
In this section, we show how highly non-Gaussian and non-classical states can emerge from a quantum control correlated with the order in which Gaussian unitary operations, i.e., Displacement and squeezing operators, are performed on an initially Gaussian state. Without loss of generality, this state is chosen to be the single mode vacuum state.

\subsection{The indefinite order of Gaussian unitaries}
Let us consider  $\hat{\rho}_G=\ketbra00_G$ be the vacuum state of a single bosonic mode, being coupled initially to the degrees of freedom of a control qubit $\hat{\rho}_C=\ketbra\psi\psi_C$ in the form $\hat{\rho}_G\otimes\hat{\rho}$. The state of the control qubit is written explicitly in the computational basis as
\begin{equation}
    \ket\psi=\frac{1}{\sqrt{2}}(\ket0_c+\ket1_c) \label{eq:control qubit}
\end{equation}

We allow the composite system to evolve according to the unitary map $\mathcal{T}$ that characterizes the quantum switch shown in Fig.~(\ref{fig:qs}), such that
\begin{equation}
    \mathcal{T}=S(r)D(\alpha)\otimes\ketbra00+D(\alpha)S(r)\otimes\ketbra11 \label{eq: quantum switch}
\end{equation}
where $S(r)$ and $D(\alpha)$ are respectively the squeezing and the displacement operators, with parameters $\alpha=x+iy$ and $r$ taken to be real without loss of generality.
Accordingly, the final state of the system  will be given by
\begin{equation}
 \frac{1}{\sqrt{2}}\big(S(r)D(\alpha)\ket0_G\ket0_c+D(\alpha)S(r)\ket0_G\ket1_c\big)   
\end{equation}
or in the coherent basis ${\ket+_c,\ket-_c}$ where 
\begin{equation}
    \ket\pm_c=\frac{1}{\sqrt{2}}(\ket0_c\pm\ket1_c) \label{eq: coherent basis} 
\end{equation}
it is given instead by 
\begin{equation}
   \frac{1}{2}\big[\big(S(r)D(\alpha)+D(\alpha)S(r)\big)\ket0_G\ket+_c
+\big(S(r)D(\alpha)-D(\alpha)S(r)\big)\ket0_G\ket-_c\big]  \label{eq: final state} 
\end{equation}

If a measurement on the control qubit is performed on the coherent basis (\ref{eq: coherent basis}), we can see from Eq.~(\ref{eq: final state}) that two post-selected states of the CV system emerging, entailing an indefinite causal order between squeezing and displacement operators. The conditional states emerge with equal probabilities and they are given by
\begin{equation}
    \ket G_\pm=\frac{1}{\sqrt{G_\pm}}\big(S(r)D(\alpha)\pm D(\alpha)S(r)\big)\ket0_G  \label{eq: emerged states}
\end{equation}
with $G_\pm=2\big[1\pm Re(\bra0 D(\alpha)^\dagger S(r)^\dagger D(\alpha)S(r)\ket0_G)\big]$ 
being the respective normalization factors.
We expect these states to be non-Gaussian as the set of Gaussian states is non-convex, as well as to be non-classical as they are not a classical mixture of Gaussian states, which makes them lie outside the convex hull of the set of single mode Gaussian states.
\begin{figure}[t]
    \centering

    \includegraphics[width=0.4\columnwidth]{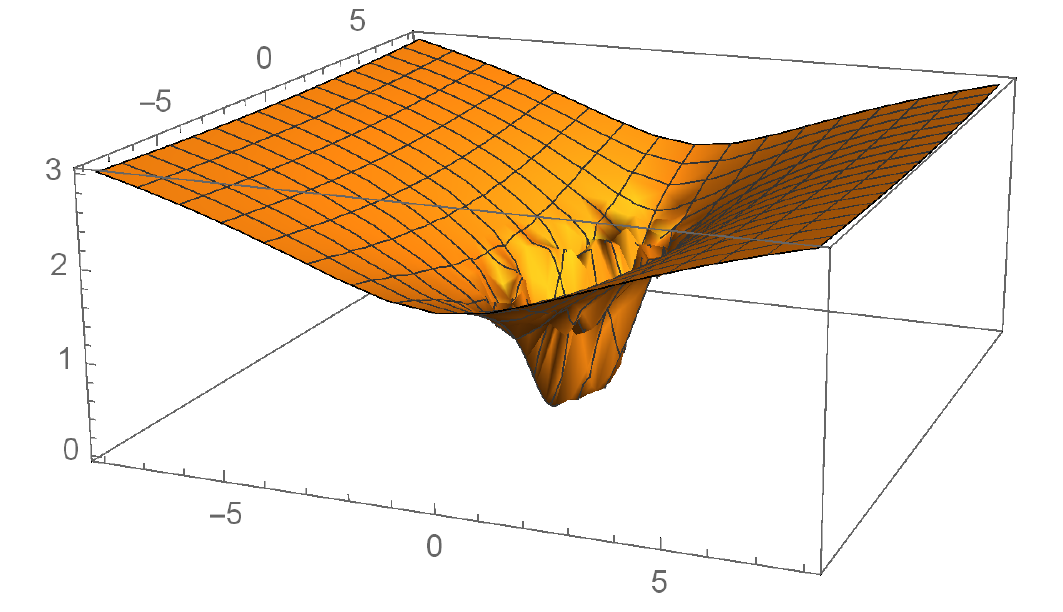}
     \includegraphics[width=0.4\columnwidth]{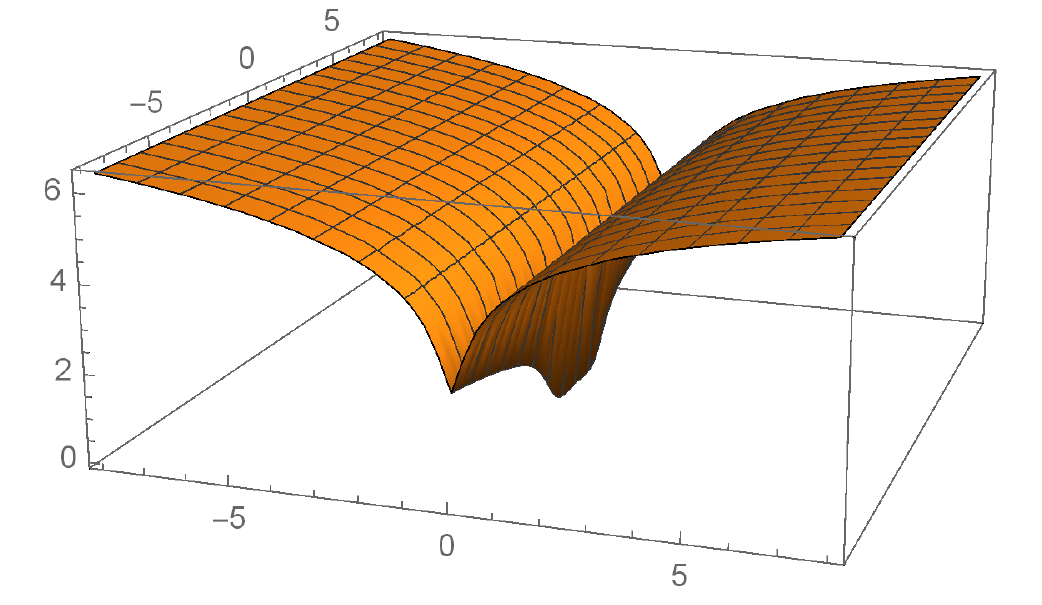}\\
     \includegraphics[width=0.4\columnwidth]{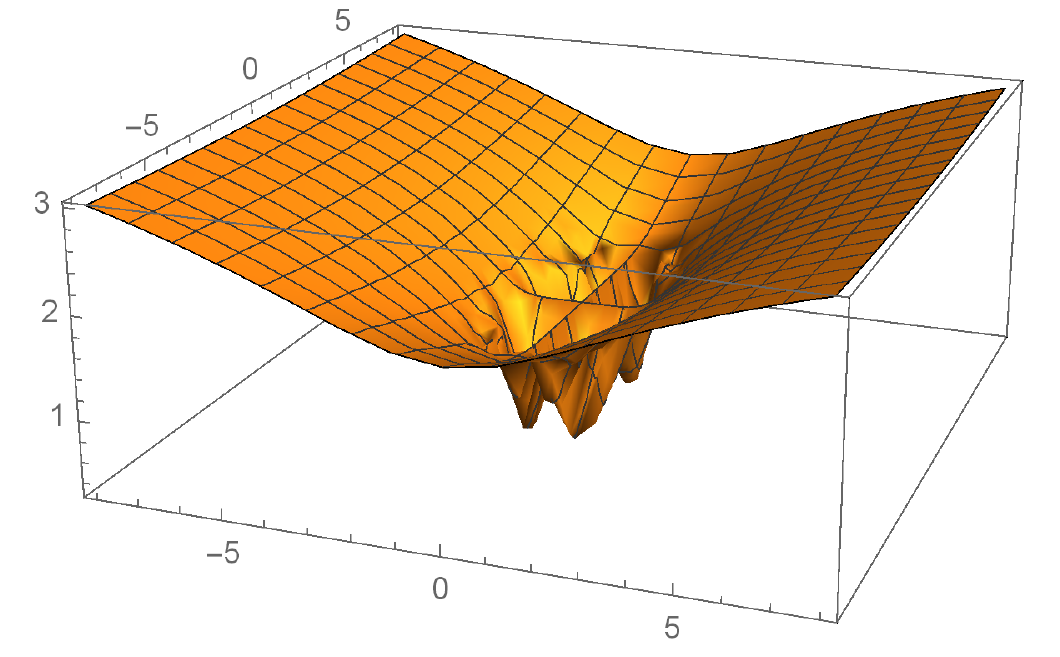}
     \includegraphics[width=0.4\columnwidth]{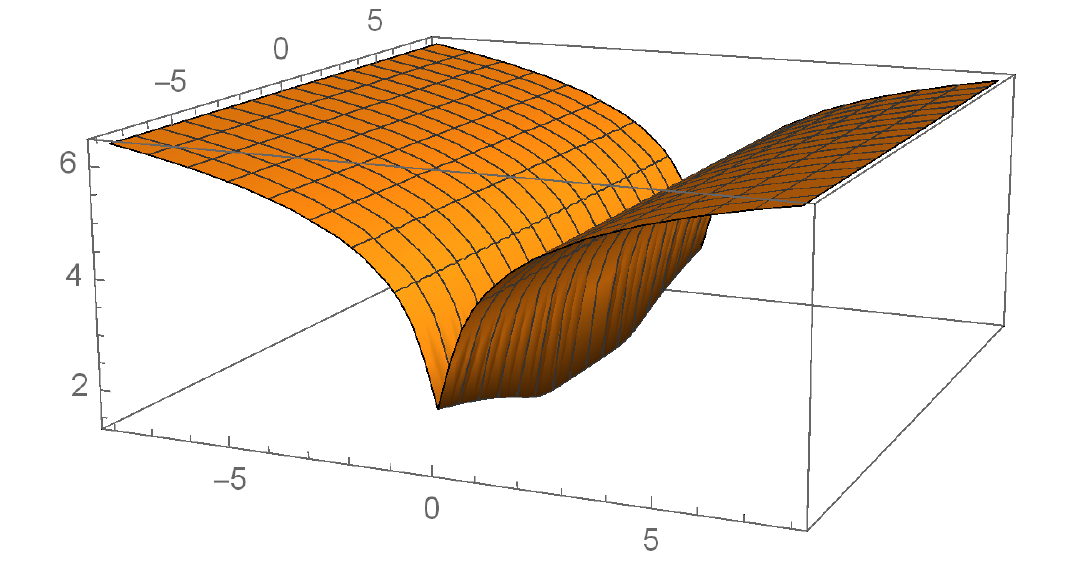}
     \caption{The quantum relative entropy $\delta_{nG}$ for the conditional states (\ref{eq: emerged states}) as a function of the displacement parameters $x$ and $y$ for different values of squeezing parameter $r$. Upper (Lower) panel represents the degree of non-Gaussianity of $\ket G_+$ ($\ket G_-$). The values of $r$ are from left to right: $r=1$ and $r=4$.   }
      \label{fig:nongaussianity}
     \end{figure}
\subsection{Properties of the resulting conditional states}
 In order to explore the non-Gaussian character of the states (\ref{eq: emerged states}) we need to determine first their covariance matrix as stated previously in sec.~(\ref{sec:2}). Using the relation between the squeezing and displacement operators 
 \begin{equation}
     S(r)D(\alpha)=D(\beta)S(r)
 \end{equation}
 for 
 \begin{equation}
     \beta=\cosh(r)\alpha+\sinh(r)\alpha^* \label{eq: beta}
 \end{equation}
 the covariance matrix elements (\ref{eq: covariance}) are thus given by 
 \begin{align}
     \textbf{V}_{11\pm}&=\frac{1}{(G_\pm)}\Big[\big(2Re(\alpha^2)+2\lvert\alpha\rvert^2+2\big)e^{-2r}-2Re(\alpha)^2
     \pm\big(2Re(K[\alpha+\beta^*]^2+1\big)e^{-2r}\nonumber\\
     & - 2e^{-2r}\big(Re(\alpha)+Re(\beta)\pm Re((\beta+\alpha^*)K)\big)^2\Big]\nonumber\\
     \textbf{V}_{22\pm}&=\frac{1}{(G_\pm)}\Big[\big(-2Re(\alpha^2)+2\lvert\alpha\rvert^2+1-2Im(\alpha)^2\big)e^{2r}\pm\big(2Re(K[\alpha-\beta^*]^2+1\big)e^{2r}\nonumber\\&-2e^{2r}\big(Im(\alpha)+Im(\beta)\pm Im((\alpha-\beta^*)K)\big)^2\Big]\nonumber\\
     \textbf{V}_{12\pm}&=\frac{1}{(G_\pm)}\Big[2Im(\alpha^2)+2Re\big(-i(\alpha^2-\beta^{*2})K\big)-2\big(Re(\alpha)+Re(\beta)\nonumber\\
     &\pm Re((\beta+\alpha^*)K)\big)\times \big(Im(\alpha)+Im(\beta)\pm Im((\alpha-\beta^*)K)\big)\big]
 \end{align}
where $K=\braket\alpha\beta=e^{-\frac{1}{2}(\lvert\alpha\rvert^2+\lvert\beta\rvert^2-2\alpha^*\beta)}$. As long as this covariance matrix is not equal to $\frac{1}{2}\mathbb{I}$, which characterizes pure Gaussian states,  $\mathbb{I}$ being the $2\times2$ identity matrix, it is a sign for non-Gaussianity. Results for the quantum relative entropy $\delta_{nG}$ of these covariance matrices, and accordingly for the conditional states (\ref{eq: emerged states}), are sketched in the plots of Fig.~(\ref{fig:nongaussianity}). We see that the conditional states emerging from the causal-nonseparability of the indefinite causal order between squeezing and displacement operations are both non-Gaussian with a significant degree of non-Gaussianity. This later being monotonic in the squeezing and displacement parameters $r$ and $\alpha$ respectively.

To further investigate the non-Classicality of the states (\ref{eq: emerged states}), and to prove that these states are outside of the single mode Gaussian convex hull, we compute their Wigner function, and according to its negativity, we deduce the non-classical features of the states. For both states the Wigner function is given by
\begin{align}
     W(q,p)&=\frac{1}{\pi G_\pm}\Big[\exp(-2v_1^2)+\exp(-2v_2^2)
    \nonumber\\
    &\pm2\exp(-2v_3^2)\cos\big[2(1-\mu)(px-qy)-2\nu(px+qy)\big]\Big]\label{eq: Wigner}
\end{align}
with 
\begin{align}
    v_1&=e^{2r}(q-x)^2+e^{-2r}(y-p)^2\nonumber\\
     v_2&=e^{2r}(q-e^{r}x)^2+e^{-2r}(e^{-r}y-p)^2\nonumber\\
     v_3&=q^2+p^2
\end{align}

\begin{figure}[t]
  \centering  
   
  \includegraphics[width=0.45\columnwidth]{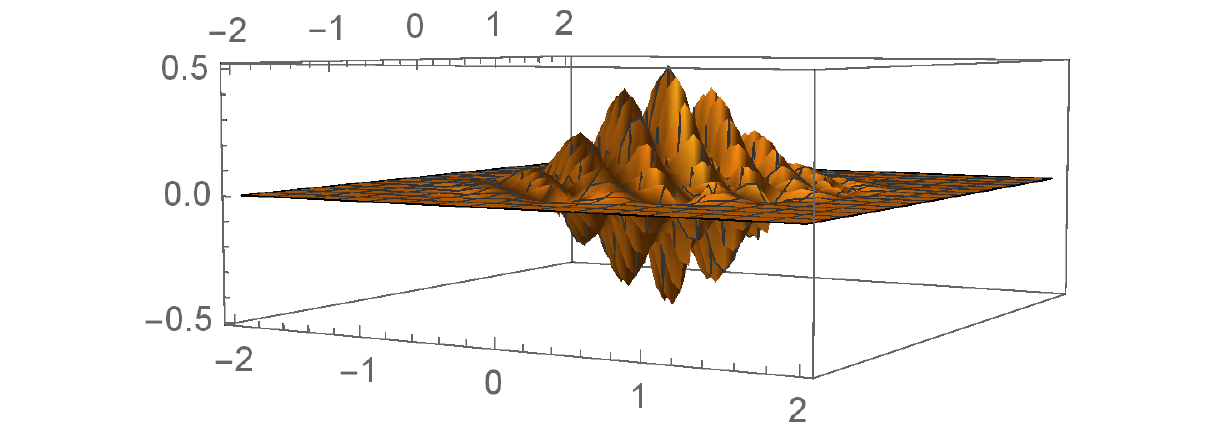}
     \includegraphics[width=0.45\columnwidth]{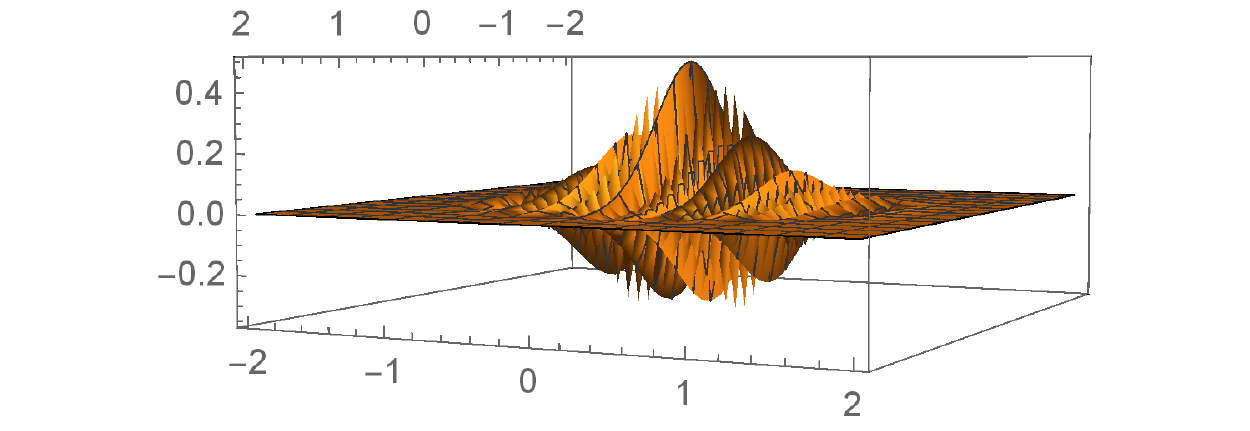}\\
    \includegraphics[width=0.45\columnwidth]{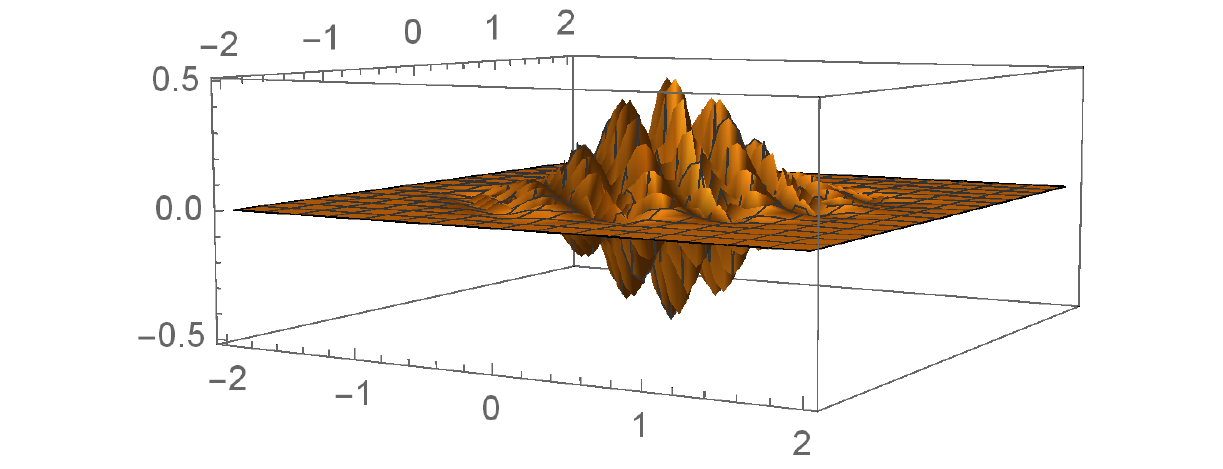}
     \includegraphics[width=0.45\columnwidth]{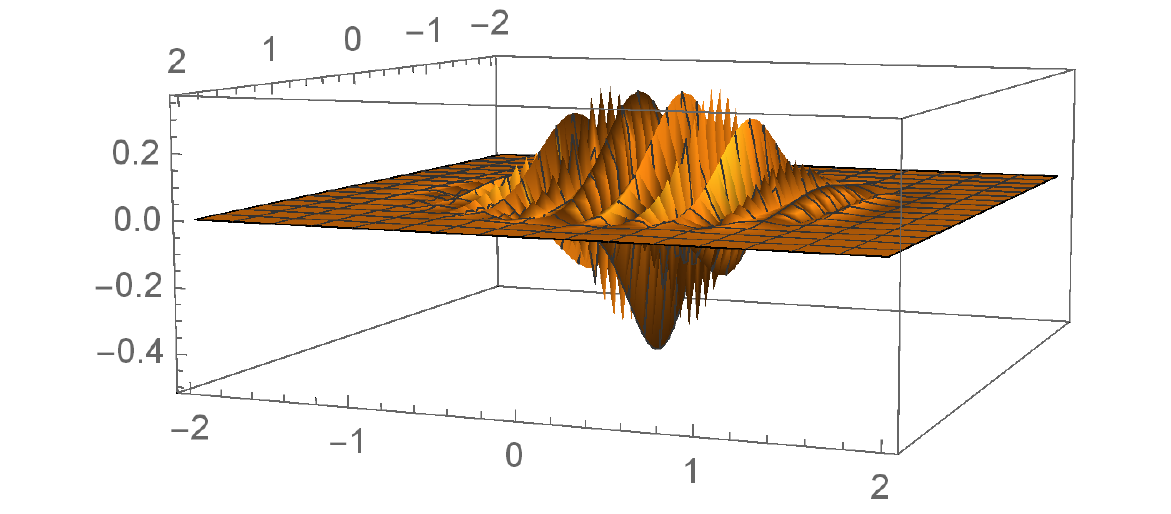}
     \caption{The Wigner function of the states (\ref{eq: emerged states}) for different values $(r,x,y)$ of squeezing parameter $r$ and displacement parameter $\alpha=x+iy$.
     Upper panel (Lower Panel): Wigner function for the state $\ket G_+$ ($\ket G_-$)  for $(1,8,0)$ and $(2,4,4)$ from left to right. }
      \label{fig:wigner}
     \end{figure}

Plots of these Wigner functions are given in Fig.~(\ref{fig:wigner}) for different values of squeezing and displacement parameters. The plots show that both post-selected states are non-classical states, as their corresponding Wigner functions are not positive, and for which the degree of the non-classical character depends on the parameters of squeezing and displacement operations.\\
These results prove the usefulness of the superposition of causal orders in a quantum switch to generate highly non-Gaussian and non-classical states from controlled Gaussian operations acting on an initial Gaussian state, where the desired level of non-Gaussianity and non-classicality of the emerging non-Gaussian state depends solely on the amount of coherence in the control qubit, as well as the values of the parameters of the Gaussian operations used in the protocol, i.e., squeezing and displacement parameters. Compared to exisiting protocols for quantum state engineering, our protocol guarantees the emergence of  non-Gaussian states in a deterministic manner, in the sense that both post-selected conditional states have the character of being non-Gaussian and non-classical, hence the probability of getting the desired state amounts to be exactly one.

\section{\label{conc}conclusion}
To summarize, we have explored the usefulness and the advantage that causal-nonseparability brings in engineering non-Gaussian and non-classical states, in the framework of the newly constructed quantum switch. In Sec.~(\ref{sec:2}) we started by giving basic properties and definitions for CV systems. We also presented some relevant measures for the non-Gaussian and non-classical features at the level of quantum states, namely, the quantum relative entropy-based measure and Wigner negativity as non-Gaussianity and non-classicality witnesses respectively. In Sec.~(\ref{sec 3}) we gave a summary of our results. We discussed the hybrid control protocol of how conditional states emerges from an indefinite causal order of Gaussian operations acting on an initially Gaussian state, chosen without any loss of generality to be the vacuum of a single bosonic mode. It has been illustrated how this indefinite causal ordering  is realised by post-selection on measurement performed on a control qubit that is coupled initially to the Gaussian state, with the relative order between Gaussian operations being entangled with the degrees of freedom of the quantum control. \\
Properties of the resulting conditional states have been discussed. In particular, it was shown that they have a highly non-Gaussian character depending monotonically on the Gaussian operations parameters, i.e. squeezing and displacement parameters. Additionally, non-classicality of these states has been treated in order to show that they don't belong to the Gaussian convex hull, which contains classical non-Gaussian states that might be realised as a statistical mixture of Gaussian states. This was a proof that the proposed scheme guarantees the emergence of non-Gaussian and non-classical states deterministically.\\  
Our results  stimulate an extension of the use of the quantum switch to multi-mode Gaussian states and operations, with the respective study of both the mode and particle entanglement contents of the resulting conditional non-Gaussian states. This might have a fundamental impact on deepening the understanding of the link between non-Gaussianity and indefinite causal ordering of operations in the quantum gravity regime, since both phenomena are conjectured to emerge in the Planck scale. Moreover, due to recent results on the simulation of causal non-separability using quantum controlled non-Markovian dynamics, it is still a question whether this later could be a good candidate for quantum state engineering, paving the way towards future investigation on the subject.

\bibliographystyle{ws-ijqi}
\bibliography{QnG}
\end{document}